\documentclass[aip,reprint,jap,groupedaddress]{revtex4-1}

\usepackage{graphics,color,url,hyperref}

\begin{document}

\title{
Site preference of dopant elements in Nd$_2$Fe$_{14}$B
}

\author{Munehisa~Matsumoto}
\affiliation{
Institute for Solid State Physics, University of Tokyo, Kashiwa 277-8581, JAPAN
}

\date{\today}

\begin{abstract}
{\it Ab initio} calculations of formation energy
and mixing energy for (Nd,R)$_{2}$(Fe,Co)$_{14}$B
[R = rare-earth elements other than Nd] are presented
to address the site preference of dopants
and the corresponding magnetic properties.
Contrasting trends between magnetic calculations and non-magnetic calculations
are discussed in conjunction with the nature of localized moments
in the metallic ferromagnetism at high temperatures. Implications on the optimal heat treatment
temperature to maximize the intrinsic properties
are discussed referring to the experimental Curie temperature.
\end{abstract}

\pacs{75.50.Ww, 75.10.Lp}

%

\maketitle

\section{Introduction}
Rare-earth permanent magnets are one of the key materials
for sustainable solutions of the energy problem. It takes
strong magnetization, accordingly strong coercive force that intrinsically
originate in the uni-axial magnetic anisotropy,
high Curie temperature, and the structure stability for a magnetic material
to be useful enough in practical applications of permanent magnets.
Unfortunately we often encounter a trade-off between those prerequisites.
The champion magnet fabricated on the basis of
Nd$_{2}$Fe$_{14}$B~\cite{sagawa_1984,croat_1984}
has been a remarkably successful case
where strong magnetization and anisotropy comes with robust structure,
but a drawback has been its low Curie temperature that is only half of
elemental Fe with the body-centered-cubic structure.
Most notably, recent developments along the line of 1:12 materials
including NdFe$_{12}$N~\cite{miyake_2014,hirayama_2015,JOM_2015}
and Sm(Fe,Co)$_{12}$~\cite{hirayama_2017} fabricated on a special substrate
to go beyond Nd$_{2}$Fe$_{14}$B have shown a promising route to stronger
magnetization and anisotropy
than Nd$_2$Fe$_{14}$B
in the typical temperature range $300~\mbox{K}\le T\le 450~\mbox{K}$ for practical usage
with the reasonably elevated Curie temperature up to $800~\mbox{K}$,
but at the sacrifice of bulk structure stability.

In order to resolve
such trade-off in ferromagnetic compounds for permanent magnets,
we look into the issue of structure stability
monitoring the trend of ferromagnetism in doped Nd$_{2}$Fe$_{14}$B to inspect
the possibility for gaining both. One of the most popular dopant elements is Co
following the celebrated Slater-Pauling curve as the promising route toward
gaining both of magnetization and Curie temperature with a small amount of Co.
Unfortunately it is known that Co degrades the intrinsic uni-axial
magneto-crystalline anisotropy (MCA) in the particular crystal structure of Nd$_{2}$Fe$_{14}$B~\cite{rmp_1991}
and the possible supplement can come from doped rare-earth elements. Indeed
Dy-doping had been the solution to sustain high-temperature MCA for the coercivity.
Since heavy rare earth (HRE) elements are not so abundant, alternative solutions to minimize
or even eliminate the usage of extra HRE have been in high demand recently~\cite{hono_2012}. It is hoped
that the present study would help from the solid state physics point of view
in finding a route toward the minimum usage of extra rare earth
elements by inspecting the energetics of doping rare earth elements
in conjunction with simultaneous
doping of transition metals. Possible relevance of the heat treatment temperature
in sample preparation for realizing a particular site selective doping
is discussed referring to the intrinsic Curie temperature and the robustness of localized
magnetic moments in the target materials.

In the next section
we describe our {\it ab initio} methods combining two codes,
namely, Korringa-Kohn-Rostoker (KKR) Green's function method
on the basis of coherent potential approximation (CPA) as implemented in AkaiKKR~\cite{AkaiKKR}
and {\it ab initio} structure optimization as implemented in another
open-source package OpenMX~\cite{OpenMX}.
Main results on doped Nd$_2$Fe$_{14}$B
are given in Sec.~\ref{sec::results} and their
implications on the nature of localized magnetic moments
and heat treatment temperature are discussed in Sec.~\ref{sec::discussions}.
Final section is devoted for conclusions and outlook.

\section{Methods and target materials}
\label{sec::methods}

\subsection{{\it Ab initio} calculations for formation energy}
\label{sec::FormationEnergy}
We start with looking at the structure stability of R$_2$T$_{14}$B where R=Nd, Sm, Dy, and Y
and T=Fe or Co with calculating the formation energy $\Delta E_{\rm f}$
which is defined as follows:
\begin{eqnarray}
&& \Delta E_{\rm f}[\mbox{R$_2$T$_{14}$B}] \nonumber \\
 & \equiv & E_{\rm tot}[\mbox{R$_2$T$_{14}$B}]
 - 2E_{\rm tot}[R] - 14E_{\rm tot}[\mbox{T}] - E_{\rm tot}[B]
\end{eqnarray}
Here $E_{\rm tot}[\mbox{M}]$ is the calculated energy for a given material $\mbox{M}$
yielded from {\it ab initio} structure optimization utilizing OpenMX within
the standard pseudopotential data provided therein~\cite{openmx_pseudopotential}.
Results from magnetically polarized and non-magnetic calculations are used for discussions
without taking into account the relativistic effects. Even though spin-orbit interaction is crucial
in getting MCA, the non-relativistic set up should be good enough
to address the structure stability and ferromagnetism considering their
dominating energy scales.
We take GGA-PBE exchange correlation to describe the electronic structure dominated
by ferromagnetic $3d$-electrons most correctly.
For rare-earth elements we use pseudopotentials with open-core approximation.
The basis set we took was \verb|Nd8.0_OC-s2p2d2f1|, \verb|Sm8.0_OC-s2p2d2f1|, 
\verb|Dy8.0_OC-s2p2d2f1|, 
\verb|Y8.0-s3p2d2f1|,
\verb|Fe6.0S-s2p2d1|, \verb|Co6.0S-s2p2d2f1|, and \verb|B7.0-s3p3d2|. There is
some subtlety with the partial core correction (PCC) in the open-core approximation
which is set by the following entry in the input to OpenMX 
\begin{verbatim}
<scf.pcc.opencore
  Nd  1
  Fe  0
  B   0
scf.pcc.opencore>
\end{verbatim}
in case of Nd$_2$Fe$_{14}$B.
Analogous set up is basically recycled
for all calculations involving rare-earth elements
in the present work. However, a problem in convergence
can happen with Sm$_2$Co$_{14}$B with this particular setup which can be worked around
if the calculation is done without PCC.
We do both of (a) open-core approximation with PCC and (b) without PCC.
Results from the setup in a) is basically taken whenever available.

For the calculations of R$_2$T$_{14}$B
the energy cutoff is set to be 500~Ry
and the number of k points was 64. Number of k points was sometimes increased up to 216
to ensure that calculated results with k points being 64 are already converging.
The above choice of the computational setup seems to give a reasonable precision in an acceptable computational cost
after monitoring the convergence with respect to the richness of the basis set, energy cutoff,
and the number of k points. The experimental structure for Nd$_2$Fe$_{14}$B found in Ref.~\onlinecite{rmp_1991}
is taken as the starting structure and {\it ab initio} structure optimization
has been done for R$_2$T$_{14}$B.

\begin{figure}
\scalebox{0.75}{
\includegraphics{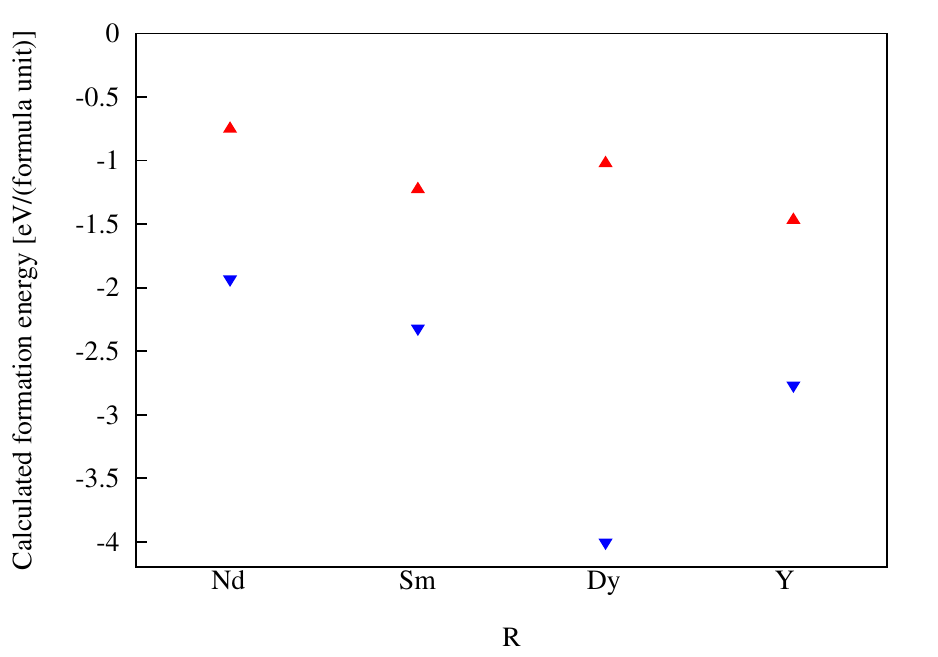}
}
\caption{\label{fig::toppage} (Color online) Calculated formation energy for pristine R$_2$T$_{14}$B (T=Fe,Co).}
\end{figure}

Energy of the reference elements is calculated in the analogous way, taking the experimental structure
of dhcp-Nd, trigonal Sm, dhcp-Dy, hcp-Y
bcc-Fe, hcp-Co and $\alpha$-B as the starting structure~\cite{MatNavi}.
The number of k points for these reference systems is taken to be 512
and the rest of the set up is the same as is done for R$_2$T$_{14}$B.
Unfortunately the structure of the elemental boron comes with polymorphism
that is not yet well controlled from first principles~\cite{ogitsu}.
The ground-state energy of Boron may not have been well fixed.
Thus the absolute values of the calculated formation energy for R$_{2}$T$_{14}$B
need be taken with some reservation
while the relative trends between materials should be able to be robustly addressed.
Thus calculated trend of the formation energy is shown in Fig.~\ref{fig::toppage}.
The details are described in Sec.~\ref{sec::sub0}.

\subsection{{\it Ab initio} calculations for mixing energy}
\label{sec::MixingEnergy}

Calculated formation energy fixes the structure stability at the stoichiometric limits
and the structure stability of off-stoichiometric compounds (Nd$_{1-x}$R$_{x}$)$_2$(Fe$_{1-y}$Co$_y$)$_{14}$B
is further addressed via mixing energy as defined in the following ways
\begin{eqnarray}
 && \Delta E_{\rm mix}[\mbox{(Nd$_{1-x}$R$_{x}$)$_2$Fe$_{14}$B}] \nonumber\\
 & \equiv & E_{\rm tot}[\mbox{(Nd$_{1-x}$R$_{x}$)$_2$Fe$_{14}$B}] \nonumber \\
 &&  - (1-x)E_{\rm tot}[\mbox{Nd$_{2}$Fe$_{14}$B}]-xE_{\rm tot}[\mbox{R$_2$Fe$_{14}$B}]
\label{eq::mix1}
\end{eqnarray}
and
\begin{eqnarray}
 && \Delta E_{\rm mix}[\mbox{Nd$_2$(Fe$_{1-y}$Co$_{y}$)$_{14}$B}] \nonumber\\
 & \equiv & E_{\rm tot}[\mbox{Nd$_2$(Fe$_{1-y}$Co$_{y}$)$_{14}$B}] \nonumber \\
 &&  - (1-y)E_{\rm tot}[\mbox{Nd$_{2}$Fe$_{14}$B}]-yE_{\rm tot}[\mbox{Nd$_2$Co$_{14}$B}].
\label{eq::mix2}
\end{eqnarray}
The trend of calculated mixing energy can point to the relative site preference of dopants.
Whenever needed, the formation energy of (Nd,R)$_2$(Fe,Co)$_{14}$B can be restored with
the following relation
\begin{eqnarray}
 && \Delta E_{\rm f}[\mbox{(Nd$_{1-x}$R$_x$)$_2$(Fe$_{1-x}$Co$_x$)$_{14}$B}] \nonumber  \\
& \equiv & E_{\rm tot}[\mbox{(Nd$_{1-x}$R$_x$)$_2$(Fe$_{1-x}$Co$_x$)$_{14}$B}] \nonumber \\
&&  - 2\left\{(1-x)E_{\rm tot}[\mbox{dhcp-Nd}]+xE_{\rm tot}[\mbox{R}]\right\} \nonumber \\
&& - 14\left[
(1-y)E_{\rm tot}[\mbox{bcc-Fe}]+0.5yE_{\rm tot}[\mbox{hcp-Co}]
\right] \nonumber \\
&& - E_{\rm tot}[B]
\label{eq::formation_energy}
\end{eqnarray}

Off-stoichiometric compounds can be continuously interpolated between stoichiometric limits
via KKR-CPA as implemented in AkaiKKR. Open-core approximation is also taken here for
rare-earth elements and the angular momentum cutoff is set to be $l_{max}=2$ which
should be sufficient in describing $3d$-electron states of Fe/Co and $5d$-electron states
coming from rare-earth elements. Also with AkaiKKR we collect the data
on the basis of non-relativistic calculations.
Within OpenMX calculations, discrete substitution of host atoms is inspected
on top of which AkaiKKR results are overlapped to ensure the consistency between two codes
and the exploration range can be extended continuously.
Mixing energy can be exploited with KKR-CPA even with systematic corrections
in calculated total energy on the basis of KKR-CPA~\cite{zeller_2013} because the corrections
in the terms on the right-hand side of analogous relations
to Eqs.~(\ref{eq::mix1})~and~(\ref{eq::mix2}) are cancelling out. The advantage
of OpenMX results is that formation energy of the stoichiometric compounds and discretely
substituted compounds can both be directly addressed while fractional substitution can be more
easily addressed via AkaiKKR. We combine these two approaches to cover relevant dopants
for possible improvement of magnetic properties of Nd$_{2}$Fe$_{14}$B. Namely,
all magnetic rare-earth elements which reasonably converged on the 2:14:1 structure
(Pr, Sm, Gd, Tb, Dy, Ho, Er, Tm, Yb$^{3+}$) to partially replace Nd and Co to partially replace Fe.
Convergence with a fictitious Eu$_2$Fe$_{14}$B seems to involve some problems and the task of
answering them is split for other study.

Effects of
non-magnetic light rare earth elements, La and Ce$^{4+}$, explicitly involve $4f$-electrons
in the valence state. This is qualitatively different from what we deal with in the present
work where $4f$ electrons are so well localized that energetics can be safely addressed
only with $3d$-electrons from Fe, $5d$-electrons from Nd, and the respective substitutes.
Energetics issue involving $4f$-electrons in (La,Ce)-based compounds
is addressed in a separate work~\cite{ito_etal}.

\section{Results}
\label{sec::results}

Calculated formation energy
of R$_2$T$_{14}$B where R=Nd, Sm, Dy, and Y
and T=Fe or Co employing OpenMX is inspected in Sec.~\ref{sec::sub0}.
In Sec.~\ref{sec::sub1} we present results for the mixing energy in Nd$_{2}$(Fe,Co)$_{14}$B
obtained with OpenMX to elucidate the $3d$-electron magnetism and energetics.
 Contrasting outcome from magnetic and non-magnetic calculations is discussed
referring to past experimental claims. Then in Sec~\ref{sec::sub2}
we discuss $5d$-electron physics in (Nd,R)$_2$Fe$_{14}$B where
R is magnetic and trivalent rare earth elements other than Nd based on both of OpenMX and AkaiKKR results
for the mixing energy. Finally in Sec.~\ref{sec::sub3}
calculated formation energy of (Nd,Dy)$_2$(Fe,Co)$_{14}$B as obtained with OpenMX
is presented.

\subsection{Formation energy of stoichiometric compounds}
\label{sec::sub0}

\begin{table}
\begin{tabular}{lcc} \hline
material & $N_{\rm atom}$ & $E_{\rm tot}~\mbox{[Hartree]}$  \\ \hline
$\alpha$-B & 36 & $-105.056$  \\ \hline
bcc-Fe & 1 & $-89.5730$  \\ \hline
hcp-Co & 2 & $-214.260$  \\ \hline
dhcp-Nd (a) & 4 & $-180.187$  \\ \hline
dhcp-Nd (b) & 4 &  $-173.765$ \\ \hline
trigonal-Sm (a) & 9 & $-418.320$  \\ \hline
trigonal-Sm (b) & 9 & $-404.194$ \\ \hline
dhcp-Dy (a) & 4 & $-197.010$  \\ \hline
dhcp-Dy (b) & 4 & $-190.898$ \\ \hline
hcp-Y & 2 & $-78.7988$  \\ \hline
\end{tabular}
\caption{\label{table::elemental_references} Calculated energy
with OpenMX on the basis of the standard pseudopotentials and the choice
of basis sets as described in Sec.~\ref{sec::methods}
for each elemental reference material.
$N_{\rm atom}$ is the number of atoms in the unit cell. For all of the data in this table,
the number of k points is $512$.}
\end{table}
\begin{table}
\begin{tabular}{lcc} \hline
material  & $E_{\rm tot}~\mbox{[Hartree/(cell)]}$ & $E_{\rm f}~\mbox{[eV/(f.u.)]}$ \\ \hline
Nd$_2$Fe$_{14}$B (a) & $-5388.247$ & $-0.76$ \\ \hline
Nd$_2$Fe$_{14}$B (b) & $-5375.386$ & $-0.65$ \\ \hline
Sm$_2$Fe$_{14}$B (a) & $-5399.783$ & $-1.24$ \\ \hline
Sm$_2$Fe$_{14}$B (b) & $-5387.206$ & $-1.10$ \\ \hline
Dy$_2$Fe$_{14}$B (a) & $-5421.933$ & $-1.03$ \\ \hline
Dy$_2$Fe$_{14}$B (b) & $-5409.683$ & $-0.86$ \\ \hline
Y$_2$Fe$_{14}$B  & $-5343.173$ & $-1.48$ \\ \hline\hline
Nd$_2$Co$_{14}$B (a) & $-6371.612$ & $-1.94$ \\ \hline
Nd$_2$Co$_{14}$B (b) & $-6358.752$ & $-1.83$ \\ \hline
Sm$_2$Co$_{14}$B (a) & (not converging) & N/A \\ \hline
Sm$_2$Co$_{14}$B (b) & $-6370.578$ & $-2.32$ \\ \hline
Dy$_2$Co$_{14}$B (a) & $-6405.562$ & $-4.01$ \\ \hline
Dy$_2$Co$_{14}$B (b) & (not converging) & N/A \\ \hline
Y$_2$Co$_{14}$B  & $-6326.555$ & $-2.77$ \\ \hline
\end{tabular}
\caption{\label{table::target_materials} Calculated energy with OpenMX
on the basis of the standard pseudopotentials and the choice
of basis sets as described in Sec.~\ref{sec::methods} and
the corresponding formation energy for each target material.}
\end{table}
Calculated energy of the reference elemental materials and
the target compounds R$_2$T$_{14}$B [R=Nd, Sm, Dy, Y and T=Fe,Co]
on the basis of OpenMX pseudopotentials and the choice
of basis as described in Sec.~\ref{sec::methods}
is summarized in Tables~\ref{table::elemental_references}
and~\ref{table::target_materials}, respectively.
The label (a) and (b) for materials involving rare-earth elements denotes explicit incorporation
of partial charge correction in open-core approximation
has been included (a) or not included (b), respectively, as described in Sec.~\ref{sec::FormationEnergy}.

Calculated trend of the formation energy is tabulated at the rightmost column
in Table~\ref{table::target_materials}
and plotted in Fig.~\ref{fig::toppage}
with the results basically
from setup (a). Only for Sm$_2$Co$_{14}$B the results with setup (b) is plotted.
Even though the structure stability of Nd$_{2}$Fe$_{14}$B is well established
experimentally, the calculated trend including Nd$_2$Fe$_{14}$B is instructive
in that the qualification of Nd$_2$Fe$_{14}$B is revealed in the middle of the overall trend
showing that Co-based compounds and heavy-rare-earth compounds are more stable energetically.
The trend with respect to rare earth should be attributed to lanthanide contraction
while the inter-relation between Fe-based series and Co-based series
reflect the competing trend between ferromagnetism and structure stability.
The magnetovolume effect expands the lattice and the intrinsic magnetization
should be well below a limit that the the given crystal structure can hold.
In this regard the particular structure of R$_2$T$_{14}$B is excellent.

Calculated formation energy for Nd$_{2}$Fe$_{14}$B amounting to
$-0.7\sim-0.8~\mbox{[eV/(formula unit)]}$
is in reasonable agreement with the previous claim~\cite{herbst_2017}
$-101~\mbox{[kJ/mol]}
 =  -1.05~\mbox{[eV/(formula unit)]}$.
As is seen in Table~\ref{table::target_materials},
the computational setup a) or b)
for the open-core approximation in OpenMX
can shift the formation energy by $0.1~\mbox{[eV/(formula unit)]}$.
Even though the ambiguity in the order of $O(0.1)~\mbox{[eV/(formula unit)]}$
remains in the absolute value of the formation energy,
the relative trends between compounds and position of dopants
should be robustly elucidated from first principles
as long as we coherently apply the setup of the calculations to all of the data
used for discussions.

\subsection{Nd$_{2}$(Fe,Co)$_{14}$B}
\label{sec::sub1}

\begin{figure}
\scalebox{0.75}{
\includegraphics{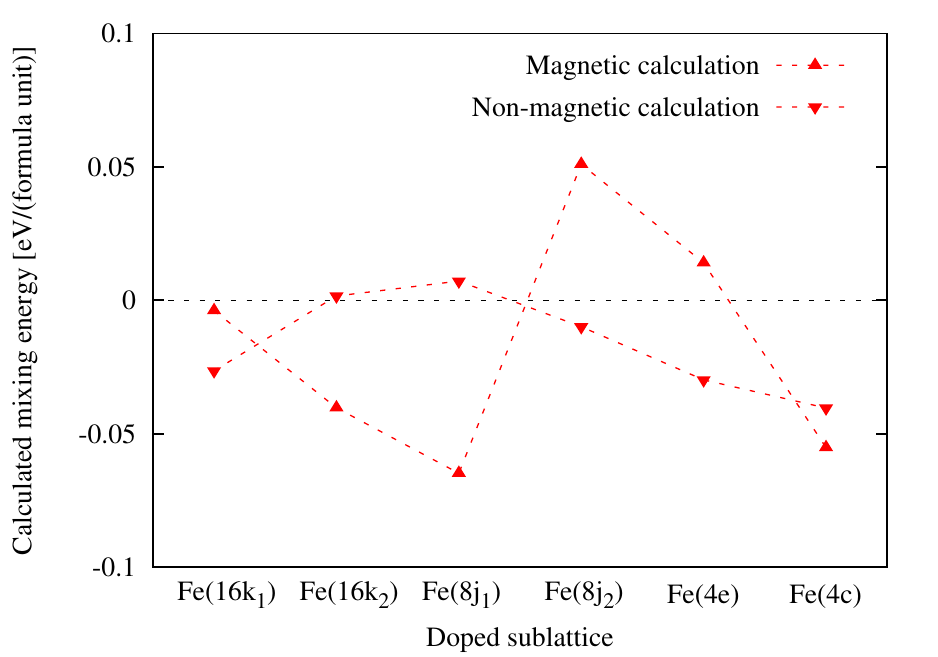}}
\caption{\label{fig::results_3d_in_Nd2FeCo14B} (Color online)
Calculated mixing energy
of Nd$_{2}$Fe$_{14}$B and Co for Co-doped Nd$_{2}$Fe$_{14}$B on the basis of discrete substitution
using OpenMX. Dotted lines have been attached to serve as guide for the eye.}
\end{figure}
Calculated mixing energy for Nd$_{2}$Fe$_{14}$B and Co in Nd$_{2}$(Fe$_{55/56}$Co$_{1/56}$)$_{14}$B is shown in 
Fig.~\ref{fig::results_3d_in_Nd2FeCo14B}. 
We note that the ambiguity in the absolute value of
calculated formation energy is cancelled out in the mixing energy.
Discrete substitution of Fe by Co is done in OpenMX
and one Co atom is put in one of the 56 possible Fe atomic sites in the unit cell. 
Magnetic calculations show that Fe($8j_1$) site is most favorable
for Co to substitute and Fe($8j_2$) site is most unfavorable, which is in remarkable agreement
with the past experimental claim measured by M\"{o}ssbauer effect~\cite{moessbauer_1993}.
The same M\"{o}ssbauer measurement~\cite{moessbauer_1993}
claimed less population in Fe($16k_1$) and some preference in Fe($16k_2$)
which is again in good agreement with the trend in calculated mixing energy on the basis of ferromagnetic
ground states while the expected preference of Fe($4c$) and avoidance of Fe($4e$) seems to be at variance
with the experimentally observed
slight avoidance on Fe($4c$) and preference for Fe($4e$)~\cite{moessbauer_1993}.

\begin{figure}
\scalebox{0.75}{
\includegraphics{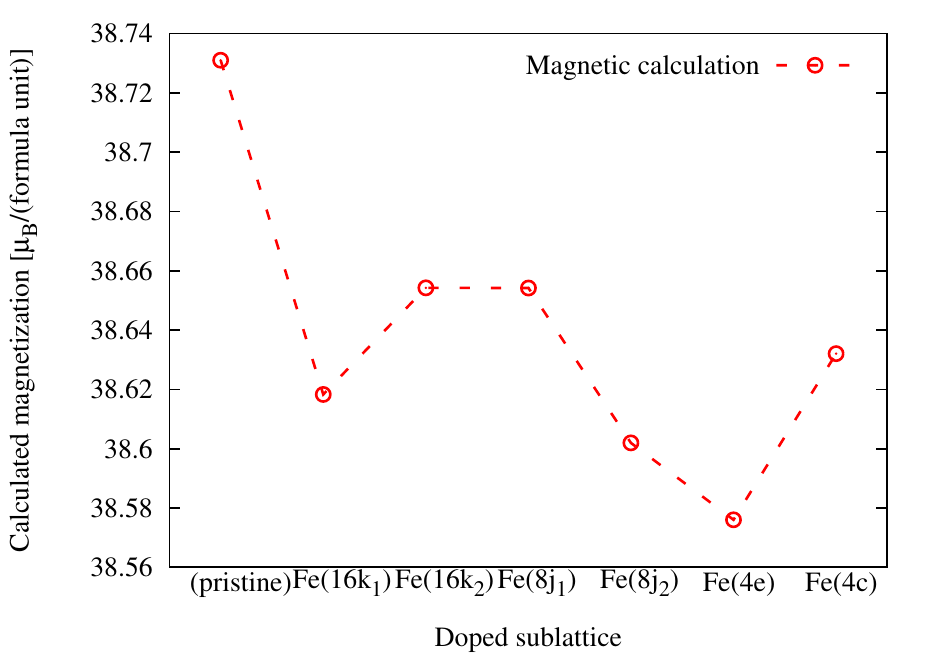}
}
\caption{\label{fig::results_mag_Nd2FeCo14B} (Color online)
Calculated magnetization
of pristine Nd$_{2}$Fe$_{14}$B and Co-doped Nd$_{2}$Fe$_{14}$B. Doping of Co has been
done on the basis of discrete substitution using OpenMX. Dotted lines have been attached to serve as guide for the eye.}
\end{figure}
Corresponding trends of magnetization is shown in Fig.~\ref{fig::results_mag_Nd2FeCo14B}. On top of open-core approximation
results, the contribution of two Nd atoms in the formula unit
has been restored with $g_J\sqrt{J(J+1)}$ for one Nd atom, where $J=9/2$ and $g_J=8/11$. It is noted
that all possible ways to dope Co into Nd$_{2}$Fe$_{14}$B lead to inferior magnetization.
This is not actually recovered by the volume effects even if the lattice shrinks upon
doping of Co. While some doping of Co lead to slightly enhanced structure
stability in Fig.~\ref{fig::results_3d_in_Nd2FeCo14B}, ferromagnetism is sacrificed here.

The calculated trend in the magnetization of Co-doped Nd$_{2}$Fe$_{14}$B is
at variance with the Slater-Pauling curve. It may well be the case
that Co-induced elevation of Curie temperature
can render the room temperature magnetization as a function of Co concentration
resembling the Slater-Pauling curve as was reported
in some of the past experimental measurements~\cite{shimotomai}.
In the ground state, we conclude
that no Slater-Pauling curve would be observed in Nd$_{2}$(Fe,Co)$_{14}$B
with the observed trend in Fig.~\ref{fig::results_mag_Nd2FeCo14B} and also
based on CPA data obtained with AkaiKKR. More systematic data for the appearance or the disappearance of Slater-Pauling curve
in $4f$-$3d$ intermetallics
will be presented elsewhere~\cite{mm_prep}.

\subsection{(Nd,R)$_{2}$Fe$_{14}$B [R$\ne$Nd]}
\label{sec::sub2}
\begin{figure}
\scalebox{0.75}{
\includegraphics{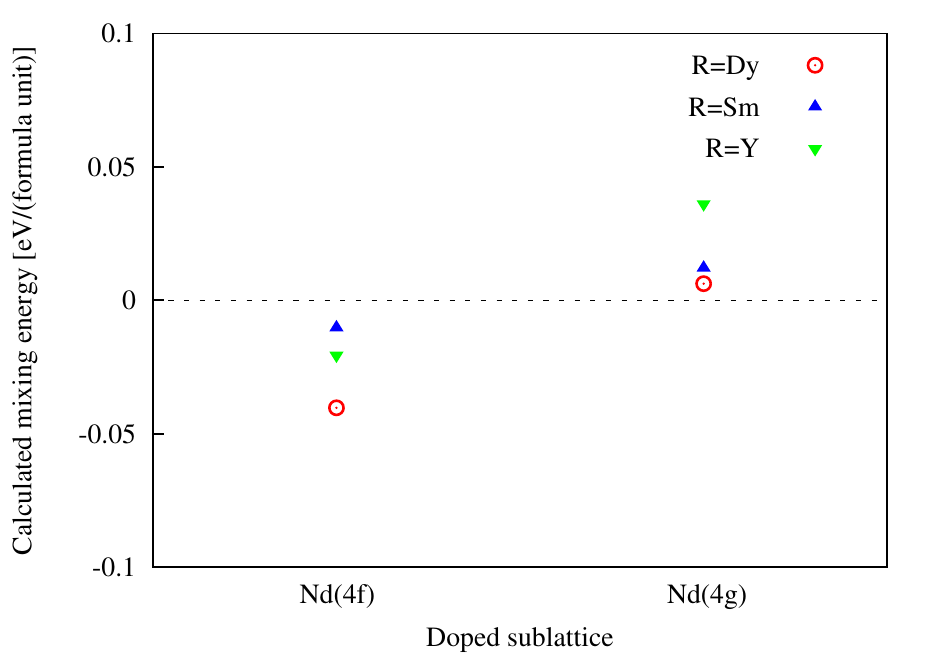}}
\caption{\label{fig::results_5d_in_NdDy2Fe14B} (Color online) Calculated mixing energy
of Nd$_{2}$Fe$_{14}$B and R for (Nd,R)$_{2}$Fe$_{14}$B (R=Dy, Sm, and Y).
Results are obtained with OpenMX on the basis of discrete substitution.}
\end{figure}

\begin{figure}
\scalebox{0.75}{
\includegraphics{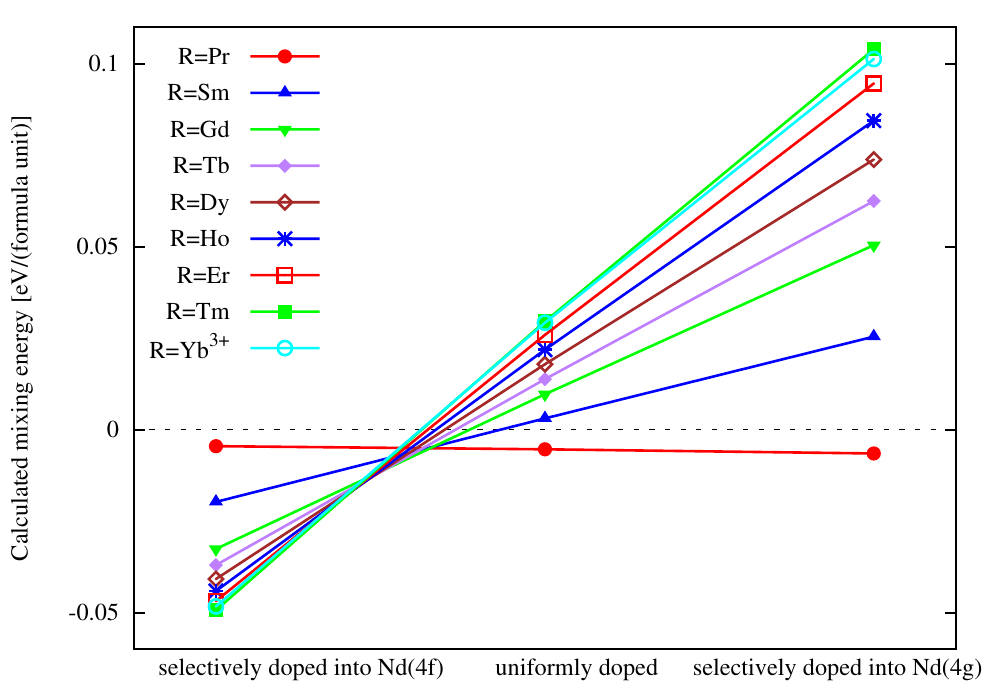}}
\caption{\label{fig::AkaiKKR_results_5d_in_NdR2Fe14B} (Color online) Calculated mixing energy
of R-doped Nd$_{2}$Fe$_{14}$B (R$\ne$ Nd). Results with continuous substitution of 10\%
on a fixed lattice of the pristine compound Nd$_{2}$Fe$_{14}$B on the basis of CPA as implemented in AkaiKKR.}
\end{figure}

Calculated mixing energy for (Nd,R)$_2$Fe$_{14}$B with R=Sm, Dy, and Y with OpenMX is shown in Fig.~\ref{fig::results_5d_in_NdDy2Fe14B}.
Discrete replacement is done in such a way
that just one dopant rare earth atom replaces one of the eight Nd atoms in the host Nd$_2$Fe$_{14}$B and {\it ab initio} structure
optimization has been done with OpenMX to yield the energy on the basis of OpenMX pseudopotentials with the setup a)
and the choice of basis as described in Sec.~\ref{sec::methods}. Preference of dopant Dy for Nd($4f$)
and calculated mixing energy is in agreement with the results in the previous work~\cite{saito_2017}.
The same thing is done on the basis of continuous replacements following KKR-CPA
for which the results obtained with AkaiKKR
are shown in Fig.~\ref{fig::AkaiKKR_results_5d_in_NdR2Fe14B} which is a plain extension of what was done
for Dy in Ref.~\onlinecite{saito_2017} toward other rare-earth elements. In the latter 10\% of the host Nd atoms
are replaced either selectively Nd($4f$)/Nd($4g$) sublattice or uniformly on a fixed lattice of
Nd$_2$Fe$_{14}$B that is known from past experiments~\cite{rmp_1991}. The results for R=Sm and Dy look reasonably consistent
between OpenMX and AkaiKKR considering the similar replacement ratio 1/8 in the former and 10\% in the latter,
and also the expected minor effects of lattice variation in the small concentration range of dopants.
With AkaiKKR we have basically covered all relevant rare-earth elements and see that except for Pr
all dopant rare earth prefers Nd($4f$) site. Here we have verified
that the difference between magnetic and non-magnetic calculations is minor.

\subsection{(Nd,Dy)$_{2}$(Fe,Co)$_{14}$B}
\label{sec::sub3}

\begin{figure}
\scalebox{0.75}{
\includegraphics{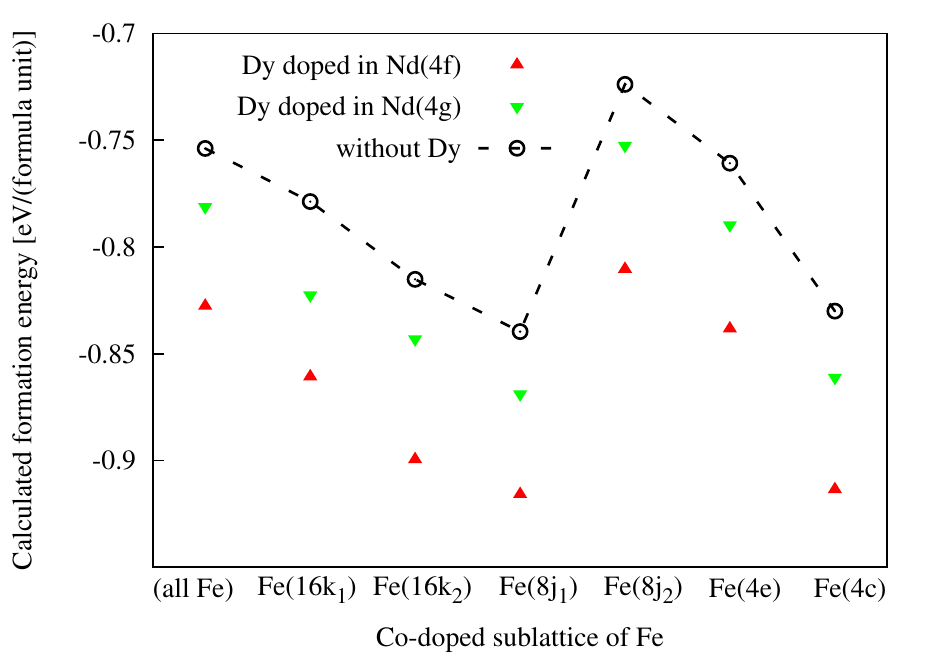}
}
\caption{\label{fig::results_5d_in_NdDy2FeCo14B} (Color online) Calculated formation energy
of (Nd,Dy)$_{2}$(Fe,Co)$_{14}$B. Results are obtained with OpenMX on the basis of discrete substitution.}
\end{figure}
Calculated formation energy for (Dy,Co)-doped Nd$_2$Fe$_{14}$B is shown in Fig.~\ref{fig::results_5d_in_NdDy2FeCo14B}.
All of the data is taken with setup a) as described in Sec.~\ref{sec::FormationEnergy}.
Discrete doping OpenMX is done with 1/8 of Nd and 1/56 of Fe being replaced by Dy and Co, respectively.
For the convenience of comparison, calculated formation energy from the magnetic calculations
as presented in Fig.~\ref{fig::results_3d_in_Nd2FeCo14B} is included here.
The overall trend in Co-position dependence of the calculated formation energy in (Nd,Dy)$_{2}$(Fe,Co)$_{14}$B
is basically just a shift from what is observed for Nd$_{2}$(Fe,Co)$_{14}$B.

We note that presence of Co
would affect the site distribution ratio of Dy over Nd($4f$) and Nd($4g$).
Because the energetics in the site preference of Co dominate over the energetics of Dy by a factor of two or three,
the seemingly unfavorable situation in Fig.~\ref{fig::results_5d_in_NdDy2Fe14B} where Dy resides in Nd($4g$) site
can be less unlikely when Co sufficiently populates Fe($8j_1$) site.

\section{Discussions}
\label{sec::discussions}

\subsection{Localized magnetic moments at high temperatures}

\begin{figure}
\scalebox{0.75}{
\includegraphics{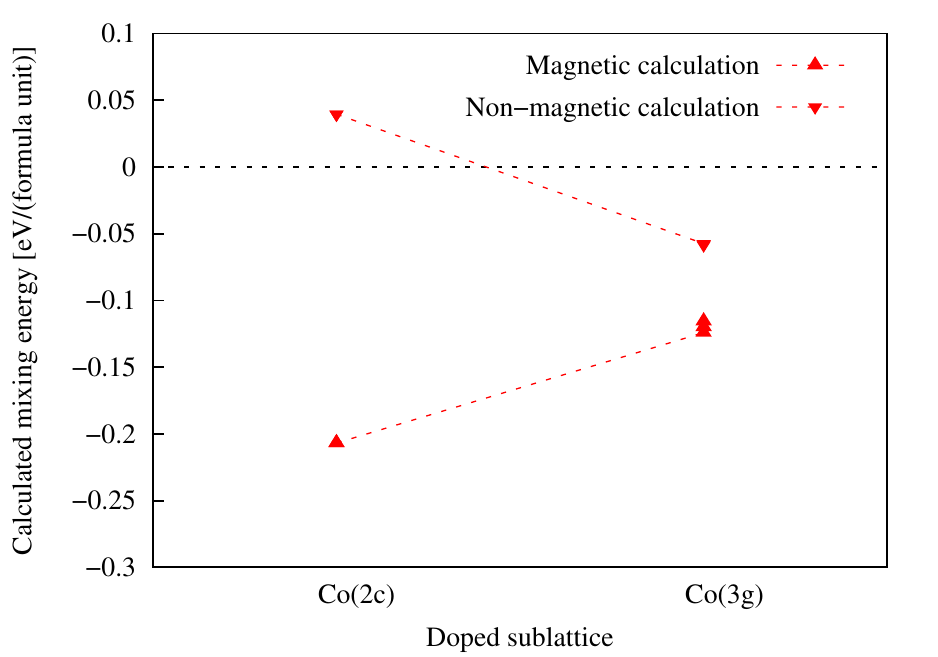}}
\caption{\label{fig::results_3d_in_RT5} (Color online) Calculated mixing energy
for Fe-doped SmCo$_5$. Results are obtained with OpenMX on the basis of discrete substitution.}
\end{figure}
We have seen that the results from magnetic and non-magnetic calculations
for Co-doped Nd$_{2}$Fe$_{14}$B show contrasting trends in Fig.~\ref{fig::results_3d_in_Nd2FeCo14B}.
Even stronger contrast is actually
seen for mixing energy calculations for Fe-doped SmCo$_5$ with the simpler crystal structure.
Calculated mixing energy for Sm(Co,Fe)$_5$
is shown in Fig.~\ref{fig::results_3d_in_RT5}. SmCo$_5$ makes the cell boundary phase
in Sm-Co based magnet which was the champion magnet in 1970's and
is still of practical importance for special-purpose permanent magnet used under extreme conditions.
The data is collected with OpenMX where we discretely replaced one Co atom by a dopant Fe atom. 
There are two sublattices, Co($2c$) and Co($3g$), and site preference of dopant Fe
points to opposite site, namely, magnetic calculations show Fe prefers Co($2c$) while
according to non-magnetic calculations Fe prefers Co($3g$) site. Result of this non-magnetic calculation 
is consistent with the past calculation~\cite{uebayashi_2002}.
Considering the high Curie temperature of SmCo$_5$ at 1000~K that can even be higher than
temperatures in sample fabrication processes,
it looks likely for SmCo$_5$ to sustain the magnetic nature
through the heat treatment processes. On the other hand the low Curie temperature
of Nd$_{2}$Fe$_{14}$B could mean that a high-temperature sample fabrication
involving Co would be more consistent
with the results of non-magnetic calculations in Fig.~\ref{fig::results_3d_in_Nd2FeCo14B}.

The contrast is seen here when some experiments on Nd$_2$(Fe,Co)$_{14}$B~\cite{moessbauer_1993} 
show remarkable agreement with our magnetic calculations for the preference of Co
among the Fe sublattices that dominate the magnetization. Even with the low Curie temperature,
high-temperature processes for the sample fabrication
preserves the message of magnetic electronic state presumably due to the robustness
of localized magnetic moments in Nd$_{2}$Fe$_{14}$B.
On the other hand, experimentally found trends for the site preference of Fe and Co in 1:5 materials
seem to point to the possibility that Fe prefers Co($3g$) site~\cite{moessbauer_1993},
being consistent rather with the above data from non-magnetic calculations. This presumably
reflects the fragility of localized magnetic moments at high temperatures in Co-based 1:5 materials.
Low (high) Curie temperature comes from relatively weak (strong) exchange couplings
among the robust (fragile) localized magnetic moments in Nd$_{2}$Fe$_{14}$B (SmCo$_5$)
at high temperatures, respectively. It seems that
robustness of localized magnetic moments and strength of exchange couplings are traded off.
{\it Ab initio} quantification of
this scenario on the basis of local moment disorder (LMD) picture
for finite-temperature magnetism~\cite{oguchi_1983,pindor_1983,
gyorffy_1985, julie_1985, julie_1986} is desired.

\subsection{Implications on the heat treatment temperature}

The messages of magnetic calculations and non-magnetic calculations for the site preference
correspond to low-temperature limit and high-temperature limit of the sample fabrication processes,
respectively. It is reasonable to see in the literature
some experimental claims
which point to various cases of the site preference of Co in Nd$_2$(Fe,Co)$_{14}$B~\cite{rmp_1991}
considering the relatively low Curie temperature of the host material Nd$_2$Fe$_{14}$B.
It is expected that the electronic state of Nd$_2$Fe$_{14}$B may not be quite
ferromagnetic in the middle of
very high temperature process when the sample is prepared. The precise boundary on the temperature axis
between the non-magnetic region and the paramagnetic region with the localized magnetic moments fluctuating
in the heat bath is yet to be explored with LMD. When the experimental
processing temperature is only slightly above the Curie temperature at 585~K,
LMD calculations would be most realistic.
Closeness of ferromagnetic results or non-magnetic results
to the experimental reality depends on the processing temperature.
If the sample is fabricated at very high temperatures and quenched, the site preference
found in the resultant sample might well be consistent with the results from non-magnetic calculations.
On the other hand, site preference in well annealed sample at relatively
low temperature should reflect the results from ferromagnetic calculations.
Heat treatment protocols for Sm-Co magnets has been extensively studied
and good chemical composition in the cell boundary phase for the permanent-magnet utility
has been nailed down~\cite{navid_2017}
and the corresponding {\it ab initio} inspection of the intrinsic magnetism
at finite temperatures has been carried out~\cite{chris_2018}.
For the 1:5 compounds in the cell boundary phase of Sm-Co magnets, such effects
of annealing temperature were discussed~\cite{gabay_2005}. We expect in more widely
applicable context that
those sample fabrication processes which are mostly off-equilibrium
can in principle be improved or even optimized
referring to these intrinsic properties at thermal equilibrium to better control
the chemical composition and site preference.

\section{Conclusions and outlook}
\label{sec::conclusions}

We have presented {\it ab initio} results
of the formation energy and mixing energy
for doped Nd$_2$Fe$_{14}$B. Revealed trends in the site preference
compared to experimental observation point to the validity of
localized magnetic moment picture for Nd$_2$Fe$_{14}$B at high temperatures
while fragile nature in the localized moments in SmCo$_5$ at high temperatures
seems to have been indicated. This may be in contrast to the trends
in the robustness of localized magnetic moments in the ground state~\cite{mm_2018}
and extra care must be taken in describing the finite-temperature magnetism.
Various trade-off situations have been elucidated between
structure stability and magnetization, or between high Curie temperature and robustness in localized magnetic moments.
At the moment the best we can do is to identify a good compromise and we have provided {\it ab initio} data
that can be of potential use for that.

Referring to the outcome of magnetic and non-magnetic calculations,
the heat-treatment temperature in the sample fabrication process
compared to the intrinsic Curie temperature can tell which site preference would be realized
at least energetically. Two extreme cases, so to say high-temperature limit
and low-temperature limit, have been inspected in the present study
while the truth for finite-temperature magnetism and structure stability
lies somewhere in between - {\it ab initio}
studies based on the LMD picture for finite-temperature magnetism
of Fe-based ferromagnets~\cite{pindor_1983,
gyorffy_1985, julie_1985, julie_1986} would be desirable here.

\begin{acknowledgments}
MM is supported by Toyota Motor Corporation.
Helpful comments given by T.~Ishikawa concerning the practical precision
of calculated formation energy and useful
discussions with M.~Morishita, T.~Abe, H.~Akai, S.~Doi,
A.~Marmodoro, M.~Hoffmann, A.~Ernst, C.~E.~Patrick, and J.~B.~Staunton in related projects
are gratefully acknowledged. MM benefited from
interactions with  Y.~Harashima, T.~Miyake, and T.~Ozaki
for {\it ab initio} calculations employing OpenMX.
The present work was partly supported by
the Elements Strategy Initiative Project under the auspice of Ministry of Education,
Culture, Sports, Science and Technology.
\end{acknowledgments}

\end{document}